\documentclass[12pt,english]{revtex4}

\usepackage{amsmath}

\newcommand{\be}{\begin{equation}}
\newcommand{\ee}{\end{equation}}
\newcommand{\bp}{\begin{pmatrix}}
\newcommand{\ep}{\end{pmatrix}}

\begin{document}

\title{From the Boltzmann equation to fluid mechanics on a manifold}
\author{Peter J. Love, Donato Cianci
{\it Haverford College}\\
{\it Department of Physics}\\
{\it 370 Lancaster Ave}\\
{\it Haverford, PA 19041}\\
}

\begin{abstract}
We apply the Chapman-Enskog procedure to derive hydrodynamic equations on an arbitrary surface from the Boltzmann equation on the surface.
\end{abstract}

\maketitle

\section{Introduction}
One of Boltzmann's greatest contributions to statistical mechanics is his kinetic equation for gases~\cite{cercignani}. The Chapman-Enskog (CE) procedure was first used  to obtain the Navier-Stokes equations from the Boltzmann equation and later adapted to analyze the macrodynamics of lattice gases~\cite{Chapman,Frisch}. Recent work has extended the lattice gas automaton to simulate fluids on arbitrary surfaces \cite{Klales}, calling for methods of analyzing lattice gas methods defined on an arbitrary surface. A first step towards such methods is the extension of the CE expansion to a curved background space. In this paper we consider Boltzmann's equation on a curved surface, and use the CE expansion to obtain hydrodynamic equations on such surfaces. 

We consider the Boltzmann equation with the Bhatnagar, Gross and Krook (BGK) collision operator~\cite{BGK}:
\begin{equation}\label{BGK}
\partial_t f + {\bf v}\cdot \nabla f = - \frac{f - f^{(0)}}{\tau},
\end{equation}
where $f^{(0)}$ is the local equilibrium distribution, $f$ is the single particle distribution function, and $\tau$ is the typical time scale of the relaxation to local equilibrium. 

 \section{The moment equations}
 
The fluid density, fluid velocity, and energy are defined in terms of $f$ by averaging over all  molecular velocities.
\be\label{macro}
\rho =  \int mf d{\bf v}, ~~\rho {\bf u} =  \int {\bf v} mf d{\bf v}, ~~~\rho {\cal E} =  \int \frac{1}{2}m|{\bf v}|^2 f d{\bf v}.
\ee
To hydrodynamic equations from~(\ref{BGK}) we multiply by the relevant moment and integrate over velocity. Using conservation of mass, momentum and energy  in collisions gives:
\begin{eqnarray}\label{macroeqns2}
&&\partial_t \rho + \nabla \cdot(\rho {\bf u})= 0\nonumber\\
&&\partial_t (\rho{\bf u}) + \nabla \cdot P= 0\nonumber\\
&&\partial_t (\rho{\cal E}) + \nabla \cdot {\bf q}= 0.
\end{eqnarray}
The pressure tensor $P$ and heat flux vector ${\bf q}$ are defined by:
\be\label{pthf}
P=\int m{\bf vv} f d{\bf v},~~~~{\bf q} = \int \frac{1}{2}m|{\bf v}|^2{\bf v} f d{\bf v}
\ee
The CE procedure enables one to derive constitutive equations giving ${\bf P}$, and ${\bf q}$ in terms of $\rho$, ${\bf u}$, and $T$ for the fluid from successive approximations to Boltzmann's equation. 

 \section{The Chapman-Enskog Expansion}

The CE expansion is based on a physical picture of relaxation to equilibrium in which the fluid rapidly relaxes to a local equilibrium $f^{(0)}$ which depends upon space and time only through the variation of its moments. These moments then relax by hydrodynamic processes. The starting point is therefore an expansion of $f$ around $f^{(0)}$:
\be\label{scales}
 f = f^{(0)} + \epsilon f^{(1)} + \epsilon^2 f^{(2)} + \dots
\ee 
The terms $f^{(k)}$ are successive correction terms (for $k>0$) and $\epsilon$ is a formal parameter used to keep track of the order of the approximation. 
 
We assume that the macroscopic variables $\rho$, ${\bf u}$, and $T$ are given by the local equilibrium distribution $f^{(0)}$. This implies:
\begin{eqnarray}\label{constraint}
 \int mf^{(k)} d{\bf v} &=0,~~~~\int mf^{(k)} {\bf v} d{\bf v} =0,~~~~\int f^{(k)} \frac{1}{2}m|{\bf v}|^2 d{\bf v} =0,
\end{eqnarray}
for $k \geq 1$. Note that for the BGK equation this is equivalent to the requirement that the collision operator conserve mass, momentum and energy.

The formal solution of the BGK equation may be written in terms of the linear differential operator  ${\cal D} = \partial_t + {\bf v} \cdot \nabla $ as
$f = (1+ \tau{\cal D})^{-1} f^{(0)}$. We break up the operator ${\cal D}$ into successive approximating operators ${\cal D}^k$:
\be
{\cal D} = \sum_{k = 1}^{\infty} \epsilon^n {\cal D}^{(k)} = \sum_{k = 1}^{\infty} \epsilon^n \left( \partial_t^{(k)} + {\bf v} \cdot \nabla \right).
\ee
Where the derivatives $\partial_t^{(k)}$ correspond to a hierarchy of time scales, with rapid relaxation to local equilibria on the shortest time scales, followed by hydrodynamic relaxation processes on longer timescales.
 
Taylor expanding $(1+x)^{-1}$ we obtain a formal solution of the differential equation:
\be
f = \left[ 1+ \sum_{k = 1}^{\infty} (-1)^k \tau^k \left(\sum_{i = 1}^{\infty} \epsilon^i {\cal D}^{(i)} \right)^k \right] f^{(0)}.
\ee
Equating this to the expansion of the distribution function~(\ref{scales}) and collecting powers of $\epsilon$ gives at first order $f^{(1)} = - \tau {\cal D}^{(1)} f^{(0)}$. We compute the first order corrections to the equilibrium distribution and the viscous corrections to the momentum tensor and heat flux that they imply. 

\section{Constitutive equations}

The corrections to the pressure tensor and heat flux at $k$th order are defined as follows:
\be
P^{(k)} = \int m{\bf vv} f^{(k)}d{\bf v},~~~~~~{\bf q}^{(k)}= \int \frac{1}{2}m|{\bf v}|^2 {\bf v}f^{(k)}d{\bf v}
\ee

To complete the derivation of the fluid equations we introduce an explicit form for the local equilibrium distribution function. We work in $D$ dimensions to make the role of dimensionality explicit.  We assume a Maxwell-Boltzmann equilibrium, although it is sufficient to assume that the local equilibrium distribution is Gallilean invariant and isotropic~\cite{Love}.
\be
f^{(0)}= \frac{\rho}{m} \left(\frac{m}{2\pi k T} \right)^{D/2} \exp\left[{\frac{-m |{\bf v} - {\bf u}|^2}{2kT}}\right].
\ee
The zeroth order energy density is:
\begin{eqnarray}
\rho {\cal E} &=&\int \frac{m}{2}|{\bf v}|^2f^{(0)} d{\bf v}= \frac{DkT\rho}{2m} +\frac{1}{2}\rho|{\bf u}|^2.
\end{eqnarray}
and the pressure tensor is:
\be\label{invP}
P^{(0)}= \int m{\bf vv} f^{(0)} d{\bf v} = \frac{\rho kT}{m} {\bf 1} + \rho{\bf u}{\bf u}= p {\bf 1} + \rho{\bf u}{\bf u}
\ee
where $p = \frac{\rho kT}{m}$ is the scalar pressure and hence the gas has an ideal equation of state. The heat flux is:
\be
{\bf q}^{(0)} = \int \frac{m|{\bf v}|^2}{2}{\bf v} f^{(0)} d{\bf v}=\rho {\bf u}\left[\frac{(D+2)kT}{2m} + \frac{1}{2}|{\bf u}|^2\right]
\ee
We will also need the heat flux tensor:
\be
Q^{(0)}=\int m{\bf vvv} f^{(0)} d{\bf v} = \frac{\rho kT}{m}\Omega\cdot{\bf u} +\rho {\bf uuu}
\ee
where $\Omega_{ijkl} = \delta_{ij}\delta_{kl}+ \delta_{ik}\delta_{jl}+ \delta_{il}\delta_{jk}$ is an isotropic fourth rank tensor.
\subsection{First Order: Euler's equations}

Defining the zeroth order terms for the pressure tensor $ P^{(0)}$ and the heat flux ${\bf q}^{(0)}$ we obtain Euler's equations for an ideal fluid:
\begin{eqnarray}
 \partial_t^{(1)} \rho +\nabla \cdot (\rho {\bf u}) =& 0 \\
 \partial_t^{(1)} (\rho {\bf u}) + \nabla \cdot P^{(0)} = & 0 \\
 \partial_t^{(1)} {\cal E} + \nabla \cdot (\rho {\bf q}^{(0)}) =& 0. 
\end{eqnarray}
Simplifying and writing the momentum and energy equations in closed form in terms of the hydrodynamic fields gives:
\be
\partial_t {\bf u} +{\bf u}\cdot\nabla {\bf u} = -\frac{\nabla p}{\rho},~~~~\partial_t T+{\bf u}\cdot\nabla T = T\nabla \cdot{\bf u}
\ee

\subsection{Second Order}

At second order we define $P^{(1)}$ and obtain the Navier-Stokes equations \cite{Wolf}. 
\begin{eqnarray}\nonumber
P^{(1)} &=& -\tau\partial_{t_1}\int f^{(0)}m{\bf vv}d^Dv-\tau\nabla \cdot\int f^{(0)}m{\bf vvv}d^Dv\nonumber\\
&=&-\tau\partial_{t_1}  P^{(0)}-\tau\nabla\cdot Q^{(0)}\nonumber\\
\end{eqnarray}
eliminating the derivatives in favor of the hydrodynamic fields using the first order equations gives:
\be
P^{(1)}=-\frac{\tau kT\rho}{m}\left\{\nabla\left({\bf u}\right) +\left[\nabla\left( {\bf u}\right)\right]^T\right\}
\ee
which is the correct form for the viscous stress in a Newtonian fluid with viscosity $\eta = \tau kT\rho/m$. As the goal of the present paper is the derivation of incompressible fluid equations we do not compute the corresponding correction to the heat flux.

\section{Differential geometry on two-dimensional surfaces}\label{diffgeo}

In order to make this paper more self-contained, and to fix notation, we briefly review some relevant aspects of differential geometry. We consider an arbitrary surface, parametrized by coordinates $a^i$ for $i=1,2$. Coordinates $x^i(a^1,a^2)$, $i \in \lbrace 1,2,3 \rbrace$ give the embedding of the surface in three-dimensional Euclidean space. The metric tensor on the surface, $g_{ij}$, is defined as: 
\be
g_{ij} = \sum_{l=1}^3 \frac{\partial x^l}{\partial a^i} \frac{\partial x^l}{\partial a^j}.
\ee
we define $g$ to be the determinant of these components of the metric tensor, in terms of which we can express the Jacobian as $J = \sqrt{g}$. If ${\bf e}_l$ are the basis vectors for the surface coordinates $u^l$ the Christoffel symbols may be defined in terms of the derivatives of these basis vectors with respect to the surface coordinates:
\be
\Gamma^i_{k\ell} = {\bf e}^i\cdot\partial_\ell {\bf e}_k=\Gamma^i_{k\ell}=\frac{1}{2}g^{im} \left(\partial_l g_{mk} + \partial_k g_{m\ell} - \partial_m g_{k\ell}\right)
\ee
A contraction of the Christoffel symbol may be related to the determinant of the metric by $\Gamma^i_{ij} = \frac{1}{2} \partial_k \log(g)$ 
and hence to the Jacobian $J \Gamma^i_{ij} = \partial_k J$.

\section{Moment equations on an Arbitrary Surface}
\label{general}

We consider the form of Boltzmann's equation on an arbitrary manifold. In the terminology of~\cite{Aris}, $f$ is a relative scalar. To obtain an absolute scalar, in terms of which we may write an invariant form of the Boltzmann equation, we must multiply by the Jacobian of the coordinate system in which $f$ is expressed. This reflects the physical fact that it is the number of particles in an elementary volume which is a scalar quantity, and if the volume form on a surface (or in an arbitrary coordinate system) varies from place to place, so must $f$, even for a constant density of particles.  The invariant form of the Boltzmann equation is therefore:
\be\label{BEInv}
\partial_t (J f)  +  v^i \partial_i (fJ) = 0.
\ee

The macroscopic quantities $\rho$ and $\epsilon$ are also densities, and therefore relative scalars, while $\rho{\bf u}$ is a relative vector. Their definitions are unchanged from those given in equations~(\ref{macro}). It should be noted that for any surface, the velocity vectors lie in a Euclidean tangent space and so we do not need to consider the variation of the volume form for the integrals over particle velocity which define the macroscopic quantities. Finally, we will consider throughout a surface which does not change as a function of time and so $\partial_t J=0$ and $\partial_t g^{ij}=0$.

\subsection{The continuity equation}

The first moment equation is:
\be
\int m \partial_t (J f) d{\bf v} + \int v^i \partial_i (m fJ) d{\bf v} = 0.
\ee
Expanding terms gives us: 
\be\label{thing1}
J \partial_t \rho + J \partial_i (\rho u^i) + \partial_i J (\rho u^i) = 0.
\ee
The results of Section~\ref{diffgeo} allows us to cast equation (\ref{thing1}) as:
\begin{equation}
\label{thing2}
 J \partial_t \rho + J \partial_i (\rho u^i) + J (\rho u^i)\Gamma^j_{ji} = 0.
\end{equation}
which is equivalent to:
\be
\partial_t \rho +  (\rho u^i)_{:i} = 0.
\ee
This is the continuity equation where the partial derivatives have been replaced by covariant derivatives on the surface.

\subsection{The momentum equation}
Multiplying~(\ref{BEInv}) by $m{\bf v}$ and integrating gives:
\be\label{momgen}
\int \partial_t (J f)m v^j d{\bf  v} +  \int v^iv^j \partial_i (Jf) d{\bf  v}= 0.
\ee
We proceed as follows. First write:
\be
\partial_t \int J fm v^j d{\bf  v} = \int \partial_t (J f)m v^j d{\bf  v} + \int (J f)m \partial_tv^j d{\bf  v}.
\ee
Usually the partial time derivative of the molecular velocity may be taken to be zero as the effects of particle interaction are taken into account only through the collision term. On an arbitrary manifold, in the absence of collisions or external forces the particles obey the geodesic equation of motion:
\be
\partial_tv^j +\Gamma^j_{ik}v^iv^k=0
\ee
and we may write:
\be
\partial_t \int J fm v^j d{\bf  v} = \int \partial_t (J f)m v^j d{\bf  v} - \int (J f)m\Gamma^j_{ik}v^iv^k  d{\bf  v}
\ee
so that the first term in~(\ref{momgen}) becomes:
\be
 \int \partial_t (J f)m v^j d{\bf  v}=\partial_t \int J fm v^j d{\bf  v} + \Gamma^j_{ik} \int (J f)mv^iv^k  d{\bf  v}.
\ee
For the spatial derivatives we proceed as for the continuity equation. We expand the partial derivative of $Jf$:
\be
\int v^iv^j \partial_i (fJ) d{\bf  v} = \int v^iv^j J\partial_i fd{\bf  v}+\int v^iv^j f\partial_i J d{\bf  v}
\ee
once more, we may the results of Section~\ref{diffgeo} to write this in terms of the Christoffel symbols:
\be
\int mv^iv^j \partial_i (fJ) d{\bf  v} = J\partial_i \int mv^iv^j fd{\bf  v}+ J\Gamma^k_{kj}\int mv^iv^j f d{\bf  v}
\ee
Hence we may write the momentum equation on a fixed manifold which does not change as a function of time:
\be
\partial_t \rho u^j+\partial_i P^{ik}+\Gamma^j_{ik} P^{ik}+\Gamma^k_{kj} P^{ij}=0.
\ee
This is Cauchy's transport equation for momentum with the divergence of the momentum tensor replaced by the corresponding covariant derivative on the surface.

\subsection{The heat equation}

Multiplying Boltzmann's equation by $\frac{1}{2}m|{\bf v}|^2 = \frac{1}{2}mg_{jk}v^jv^k$ and integrating gives:
\be\label{heat}
\int \frac{1}{2}m g_{jk}v^jv^k \partial_t (J f)d{\bf  v} +  \int \frac{1}{2}m g_{jk}v^jv^kv^i \partial_i (Jf)d{\bf  v}= 0.
\ee
again, assuming the surface is time independent so that $\partial_tJ=0$ and $\partial_t g_{ij}=0$
and using the same techniques as above for the last term here we obtain:
\be
J\partial_t (\rho {\cal E}) +  J \partial_i q^i + J\Gamma^k_{ki}q^i=0
\ee
where we have defined the heat flux vector:
\be
q^i= \int \frac{1}{2}m g_{jk}v^jv^kv^i fd{\bf  v} 
\ee
Again we recognize this as the equation for the transport of heat with partial derivatives replaced by convariant derivatives.
\be
\partial_t (\rho {\cal E}) +   q^i_{:i} =0
\ee

\section{Chapman-Enskog expansion on a manifold}

We proceed with the multiscale expansion of~(\ref{scales}), but now the expansion will be expressed in terms of derivatives on the manifold. The equilibrium distribution function is:
\be
f^{(0)}= \frac{\rho}{m} \left(\frac{m}{2\pi k T} \right)^{D/2} J \exp\left[{\frac{-m g_{ij}(v^i-u^i)(v^j-u^j)}{2kT}}\right].
\ee
the derivatives $D^{(k)}$ may now be written:
\be
\hat {\cal D}^{(k)} =\partial_t^{(k)} + v^i\partial_i +\Gamma^j_{ji}v^i
\ee
As above, we only require the first correction to the equilibrium distribution, which may be written:
\be
f^{(1)} = -\tau \hat{\cal D}^{(1)} f^{(0)}.
\ee

\subsection{First order}

At first order we may compute the integrals which define the zeroth order pressure tensor, heat flux and heat flux tensor:
\begin{eqnarray}
{P^{(0)}}^{ij} &=& \int m f^{(0)} v^iv^j d{\bf v} = p g^{ij} + \rho u^iu^j\nonumber\\
{q^{(0)}}^i&=&\frac{1}{2}g_{jk}{Q^{(0)}}^{ijk} = \rho u^i\left[\frac{(D+2)kT}{2m} + \frac{1}{2}g_{ij}u^ju^k\right]\nonumber\\
{Q^{(0)}}^{ijk} &=& \int m f^{(0)} v^iv^jv^k d{\bf v} = \frac{\rho kT}{m} (u^ig^{jk} +u^jg^{ik} +u^kg^{ij}) + \rho u^iu^ju^k.
\end{eqnarray}
Substituting these in the moment equations and simplifying for the case of an incompressible fluid with $u^i_{:i}=0$ gives the Eulers equations for the fluid:
\begin{eqnarray}
\rho\partial_t u^i  + g^{ij} \partial_j p + \rho u^j\partial_j u^i + \rho u^ju^k\Gamma^i_{jk}&=&0\nonumber\\
\frac{\partial T}{\partial t} + u^k\partial_k T = Tu^k_{:k}
\end{eqnarray}
and so the energy equation in the incompressible limit reduces to the statement that the convective derivative of temperature is zero.

\subsection{Second order}

At second order we must evaluate the correction to the stress tensor $P^{(1)}$:
\be
P^{(1)} =-\tau \int\hat {\cal D}^{(1)}f^{(0)} mv^iv^j d{\bf v} = -\tau\int mv^iv^j\biggl(\partial_t^{(1)} + v^k\partial_k +\Gamma^k_{kl}v^l\biggr)f^{(0)}d{\bf v}
\ee
once again we must correctly take into account geodesic motion when expressing $P^{(1)}$ in terms of gradients of lower-order moments. In particular:
\begin{eqnarray}
\int mv^iv^j\partial_t^{(1)}f^{(0)}d{\bf v} &=&\partial_t^{(1)} {P^{(0)}}^{ij} - \int f^{(0)}mv^j\partial_t^{(1)}v^id{\bf v} -\int f^{(0)}mv^i\partial_t^{(1)}v^jd{\bf v}\nonumber\\
&=&\partial_t^{(1)}{P^{(0)}}^{ij} + \Gamma^i_{kl} {Q^{(0)}}^{jkl} + \Gamma^j_{kl} {Q^{(0)}}^{ikl}
\end{eqnarray}
so that:
\be
{P^{(1)}}^{ij} = -\tau\biggl(\partial_t^{(1)}{P^{(0)}}^{ij}  + {Q^{(0)}}^{lij}_{:l}\biggr)
\ee
by some lengthy but straightforward manipulations using our previously obtained first order quantities and their evolution equations we obtain:
\be
{P^{(1)}}^{ij} = -\frac{\tau\rho kT}{m}\biggl[g^{lj}u^i_{:l} + g^{il}u^j_{:l}\biggr]
\ee
which is recognizable as the generalization of the Newtonian viscous stress tensor where the usual partial derivatives have been replaced by covariant derivatives.

\section{Conclusion}

We have shown that the CE expansion on a manifold gives the Navier-Stokes equations in which the partial derivatives in the stress tensor are replaced by the covariant derivatives on the manifold. These are the correct fluid equations on the manifold, but these calculations offer us some additional physical insight. The fact that the  covariant divergence of the pressure tensor has two terms involving the Christoffel symbols now has an explanation in terms of the underlying microscopic model. One of these terms arises because of variation of the volume form on the manifold. This term is also present in the divergence of the flux of scalar quantities. The second of these terms accounts for the fact that particles move on geodesics. This motivates a corresponding demonstration for the discrete fluid models on triangulations of~\cite{Klales} as follows: provided the triangulation approximates the volume form and geodesics of a manifold, the macrodynamics should approximate fluid dynamics on that manifold. We leave such a demonstration for future work. While the motivation for this work is two-dimensional manifolds relevant for~\cite{Klales} the extension to $D$ dimensional manifolds is straightforward. This work is also of interest when considering fluids in the presence of background forces and interactions. Just as the Newtonian notion of force is replaced by the idea of motion on geodesics in General Relativity, one could include the effects of an external force by a modification of the background geometry. Because the models defined in~\cite{Klales} also allow the geometry to change in response to the particle content the geodesics of the triangulation could perhaps encode the effects of the mean field forces present in an interacting fluid.

\end{document}